\documentclass[nobibnotes,nofootinbib]{revtex4}
\usepackage{graphicx,color}
\usepackage{epstopdf}
\usepackage[T1]{fontenc}
\usepackage[utf8]{inputenc}
\usepackage{amsmath,bm}
\usepackage{amsfonts}
\usepackage{amssymb}
\usepackage{makeidx}
\usepackage{longtable}
\usepackage{multirow}
\usepackage{bigstrut}
\usepackage{chngcntr}
\usepackage{array}
\usepackage{booktabs}
\usepackage{float}
\usepackage{xcolor}

\newcommand{\beq}{\begin{equation}} \newcommand{\eeq}{\end{equation}}
\newcommand{\bea}{\begin{eqnarray}} \newcommand{\eea}{\end{eqnarray}}
\newcommand{\bear}{\begin{eqnarray*}} \newcommand{\eear}{\end{eqnarray*}}

\newcommand{\xx}{\mathbf{x}}
\newcommand{\yy}{\mathbf{y}}
\newcommand{\bb}{\mathbf{b}}
\newcommand{\rr}{\mathbf{r}}
\newcommand{\qq}{\mathbf{q}}
\newcommand{\kk}{\mathbf{k}}

 \def\simge{\mathrel{%
    \rlap{\raise 0.511ex \hbox{$>$}}{\lower 0.511ex \hbox{$\sim$}}}}
\def\simle{\mathrel{
   \rlap{\raise 0.511ex \hbox{$<$}}{\lower 0.511ex \hbox{$\sim$}}}}

\usepackage{tikz}
\usetikzlibrary{arrows,shapes}
\usetikzlibrary{trees}
\usetikzlibrary{matrix,arrows} 				% For commutative diagram
\usetikzlibrary{positioning}				% For "above of=" commands
\usetikzlibrary{calc,through}				% For coordinates
\usetikzlibrary{decorations.pathreplacing}  % For curly braces
\usepackage{pgffor}							% For repeating patterns

\usetikzlibrary{decorations.pathmorphing}	% For Feynman Diagrams
\usetikzlibrary{decorations.markings}
\tikzset{
    vector/.style={decorate, decoration={snake}, draw},
	provector/.style={decorate, decoration={snake,amplitude=2.5pt}, draw},
	antivector/.style={decorate, decoration={snake,amplitude=-2.5pt}, draw},
    fermion/.style={draw=black, postaction={decorate},
        decoration={markings,mark=at position .55 with {\arrow[draw=black]{>}}}},
    fermionbar/.style={draw=black, postaction={decorate},
        decoration={markings,mark=at position .55 with {\arrow[draw=black]{<}}}},
    fermionnoarrow/.style={draw=black},
    gluon/.style={decorate, draw=black,
        decoration={coil,amplitude=4pt, segment length=5pt}},
    scalar/.style={dashed,draw=black, postaction={decorate},
        decoration={markings,mark=at position .55 with {\arrow[draw=black]{>}}}},
    scalarbar/.style={dashed,draw=black, postaction={decorate},
        decoration={markings,mark=at position .55 with {\arrow[draw=black]{<}}}},
    scalarnoarrow/.style={dashed,draw=black},
    electron/.style={draw=black, postaction={decorate},
        decoration={markings,mark=at position .55 with {\arrow[draw=black]{>}}}},
	bigvector/.style={decorate, decoration={snake,amplitude=4pt}, draw},
}

\usepackage{hyperref}
\hypersetup{colorlinks=true, linkcolor=blue, citecolor=blue, filecolor=blue, urlcolor=blue,
            pdfproducer={Latex},
            pdfcreator={pdflatex}}

\begin{document}

\title{Exclusive vector meson production with leading neutrons in a saturation
 model for the dipole amplitude in mixed space}

\author{J. T. Amaral and V. M. Becker}
\affiliation{Instituto de Matem\'{a}tica, Estat\'{\i}stica e F\'{\i}sica -- FURG, 
Rio Grande, 96203-900 Rio Grande do Sul, Brazil}

\begin{abstract}

We investigate $\rho$ vector meson production in $ep$ collisions at HERA with leading neutrons
in the dipole formalism. The interaction of the dipole and the pion is described
in a mixed-space approach, in which the dipole-pion scattering amplitude is given by
the Marquet-Peschanski-Soyez saturation model, which is based on the traveling wave
solutions of the nonlinear Balitsky-Kovchegov equation. We estimate the magnitude
of the absorption effects and compare our results with a previous analysis of the
same process in full coordinate space. In contrast with this approach,
the present study leads to absorption $K$ factors in the range of those predicted by
previous theoretical studies on semi-inclusive processes.

\end{abstract}

\maketitle

%%%%%%%%%%%%%%%%%%%%%%%%%%%%%%%%%%%%%%%%%%%%%%%%%%%%%%%%%%%%%%%%%%%%%%%%%%%%%%%%%%%%%%%%%%%%%%%%%%%%%%%%%%%%%%%%%%%%%%%%%%%%%%%%%%%%%%%%%%%%%%%%%%%%

\section{Introduction}

At high energies, the interaction between the virtual photon and a proton
is described in a convenient way: long before the interaction, the photon splits
into a quark-antiquark pair, or a dipole, and this dipole interacts with the proton. Thus,
in this picture, the virtual photon-proton ($\gamma^*p$) cross section can be written in terms of the
dipole-proton cross section, which, by its turn, is related to the (imaginary
part of the) dipole-proton scattering amplitude. The simplest processes that can be described
in terms of the dipole-proton amplitude are high energy electon-proton ($ep$) collisions, which
were performed at the Hadron Electron Ring Accelerator (HERA). These collisions can produce
a variety of interesting final states and, among them, those which contain {\it leading neutrons},
which are basically neutrons produced at very small polar angles with respect to the initial proton
beam direction, carrying a large fraction $x_L$ of its momentum.
These particles can provide us important insight on strong interactions in the soft regime and,
at HERA, have been first observed in semi-inclusive reactions ($e+p\to e+ n + X$, where
$n$ represents the neutron and $X$ is a generic hadronic final state). Currently, we have access to recent high precision data on leading neutron
production, in both semi-inclusive \cite{Aaron2010,Olsson2014,Andreev2014} and exclusive reactions \cite{Andreev2016}, but already from previous HERA measurements we have learned that their production are dominated
by pion exchange \cite{Adloff1999,Chekanov2002,Chekanov2004,Aktas2005,Chekanov2005,Chekanov2007}.
Thus, the virtual photon emitted from the incoming electron interacts with the pion (of the proton cloud) which allows us to extract the $\gamma^* \pi$ cross section. The interaction between the virtual photon
and the pion can also be described in terms of dipole-pion collision, which leads us
to deal with the dipole-pion scattering amplitude. Although our knowledge on this amplitude
is still limited, it can be related to a quantity whose behavior at high energies
is much better known, the dipole-proton scattering amplitude.

According to the effective theory of color glass condensate (CGC)
\cite{JalilianMarian,PhysRevD.59.014014,PhysRevD.59.014015,PhysRevD.62.114005,Iancu2001133,Iancu2001145,Iancu2001583,Ferreiro2002489}
for high energy quantum chromodynamics (QCD), the dipole-proton scattering amplitude, in the
limit of large number of colors ($N_c$), is the solution of the Balitsky-Kovchegov
(BK) equation \cite{Balitsky,Kovchegov}, which is a nonlinear equation which gives the evolution of this
amplitude with energy. Although being the simplest nonlinear evolution equation of
high energy QCD, BK equation has no exact solution, but analytical expressions
at asymptotical regimes can be obtained, and these can be used to construct
phenomenological models for the dipole-proton amplitude. One of these models is
the so-called bCGC model \cite{Kowalski1,Kowalski2}, which is a generalization of the
Iancu-Itakura-Munier model \cite{Iancu:2003ge}, with the introduction of the dependence on
the impact parameter of the dipole-proton collision. It is important to point out
that bCGC model is a pure coordinate-space model, which means that, besides depending on
the energy, the amplitude depends on the dipole size and on the impact parameter.
This model has been used in the description of the recent HERA data
on leading neutron production in both semi-inclusive \cite{Carvalho:2015eia} and
exclusive \cite{Goncalves:2015mbf} cases. 
In the latter case, the authors applied
bCGC model to describe the recent HERA data on exclusive $\rho$ meson with leading neutron and,
besides the good description of the data, they estimated the contribution of the absorption
corrections. 
These are related to photon rescattering after interacting with the pion emitted by
the proton and are usually included in an overall constant $K$-factor.
Theoretical studies of semi-inclusive processes predict that the $K$-factor should
not be smaller than $0.7$ \cite{Nikolaev:1997cn,Alesio}. For exclusive processes, experimental
and theoretical information on absorption corrections still lack, but in the coordinate-space phenomenological
analysis done in \cite{Goncalves:2015mbf}, the values obtained should not be larger
than 0.3, meaning that in exclusive processes the absorption effects would be stronger
than in semi-inclusive ones.

Not only coordinate-space models for the dipole-proton amplitude are available
in the literature. In particular, the saturation model proposed by Marquet, Peschanski
and Soyez (MPS model) \cite{Marquet:2007qa} gives an expression for this amplitude in mixed space, which
means that it depends on the collision energy, the dipole size and, instead of depending
on the impact parameter, it depends on the momentum transferred in the dipole-proton collision.
This model is based on a fundamental property of the BK equation: at asymptotic high energies,
it presents traveling wave solutions \cite{Munier}, which have been a natural explanation, for example,
to the observed geometric scaling at HERA \cite{Stasto}. 
These solutions were originally obtained
in the simplified case where the amplitude does not depend on the impact parameter (or, equivalently,
on the transferred momentum) \cite{Munier}, but also found to exist in mixed space
and in full momentum space (when the amplitude depends on the dipole transverse momentum
and the transferred momentum) \cite{Marquet1,Marquet2}. 
In this paper we apply the MPS model
to the description of HERA data on exclusive $\rho$ meson production with leading neutron
and compare our results with those obtained in coordinate-space using bCGC model.
As in \cite{Goncalves:2015mbf}, we assume a simple relation between the dipole-pion and
dipole-proton amplitudes and test the same models for the virtual pion momentum distribution of the
proton. Among other important differences between  both analyses, we find that the magnitude
of the absorption effects are of the same order of that predicted for semi-inclusive processes,
which is in strong contrast to the results obtained in coordinate space.

This paper is organized as follows. In Sec. \ref{sec:dipole-model} we review
the description of exclusive vector meson production with
leading neutron in the dipole formalism. Section \ref{sec:amplitude} is
devoted to the traveling wave solutions of BK equation, how these have been
generalized to the mixed-space case and the description of the MPS model.
In Sec. \ref{sec:results} we apply MPS model to the description of the
HERA data on exclusive $\rho$ meson production with leading neutron, present
our results and compare them to those obtained in full coordinate space.
We summarize our main conclusions in Sec. \ref{sec:conclusions}.

%%%%%%%%%%%%%%%%%%%%%%%%%%%%%%%%%%%%%%%%%%%%%%
\section{Exclusive vector meson production with leading neutron in the
dipole model}\label{sec:dipole-model}

In the one-pion-exchange approximation \cite{OPE}, valid for small values of the
(measured) neutron transverse momentum $p_{T,n}$, the $\gamma^*p$ cross section involving leading
neutron production can be written as
\begin{equation}\label{d2sdxdt}
\frac{d^2\sigma(W,Q^2,x_L,t)}{dx_Ldt}=f_{\pi/p}(x_L,t)\sigma_{\gamma^*\pi}
(Y,Q^2),
\end{equation}
where $W^2$ is the square of the center-of-mass energy of the virtual
photon-proton system, $Q^2$ is the photon virtuality, $Y$ is the total rapidity
interval of the photon-pion system. $x_L$, the fraction of the incoming proton beam energy
carried by the leading neutron, and $t$, the four-momentum transfer squared
at the proton vertex, related to the other kinematical variables by
\begin{equation}\label{eq:t}
t\simeq -\frac{p_{T,n}^2}{x_L}-\frac{(1-x_L)(m_n^2-m_p^2x_L)}{x_L},
\end{equation}
where $m_n$ ($m_p$) is the neutron (proton) mass.
The function $f_{\pi/p}(x_L,t)$, called pion flux or
pion splitting function, describes the splitting $p\to n\pi^+$ and basically
gives the virtual pion momentum distribution in a dressed nucleon (in our
case, the proton). 
There are some parametrizations $f_{\pi/p}(x_L,t)$ in the
literature; here we will use five of them, and their explicit forms will be
given in Sec. \ref{sec:results}.

The interaction of the virtual photon with the pion is given by the cross section $\sigma_{\gamma^*\pi}(Y,Q^2)$,
which is given
\begin{equation}\label{eq:sigma-Vpi}
\sigma_{\gamma^*\pi}(Y,Q^2)=\sum_{i=T,L}\int_{-\infty}^{0}
\frac{d\sigma_i^{\gamma^*\pi\to V\pi}}{dt^\prime}dt^\prime
=\frac{1}{16\pi}\sum_{i=T,L}\int_{-\infty}^{0}
\left|{\cal A}_i^{\gamma^*\pi\to V\pi}\right|^2
\left(1+\beta_{i}^{2}\right)R_{g}^{2}\ dt^\prime,
\end{equation}
where ${\cal A}_i^{\gamma^*\pi\to V\pi}$ is the imaginary part of the scattering
amplitude and the integration is performed over the for-momentum transfer momentum
squared at the pion vertex, $t^\prime=-\qq^2$, $\qq$ denoting the transverse momentum
transferred by the pion during the collision with the virtual photon. The factor
$\left(1+\beta_{i}^{2}\right)$ accounts for the contribution of the real part of the
amplitude, which can be obtained by the dispersion relations \cite{Kowalski1}
\begin{equation}
 \beta_{i}=\tan\left(\frac{\pi\lambda}{2}\right),
 \qquad
 \lambda=\frac{\partial\log({\cal A}_i^{\gamma^*\pi\to V\pi})}{\partial Y},
 \end{equation}
and $R_{g}^{2}$ in (\ref{eq:sigma-Vpi}) incorporates the namely skewness effect
(for more details, see \cite{Amir}). $\sigma_{\gamma^*\pi}(Y,Q^2)$ depends on the
photon virtuality and the total rapidity interval of the photon-pion system, given by
\begin{equation}\label{eq:rapidity}
Y=\ln\left(\frac{Q^2+\hat{W}^2}{Q^2+M_V^2}\right),
\end{equation}
where $\hat{W}^2=W^2(1-x_L)$ the square of the center-of-mass energy of the $\gamma^*\pi$
system and $M_V$ is the mass of the produced vector meson.

\begin{figure}[!h]
 \begin{center}
  \includegraphics[scale=0.8]{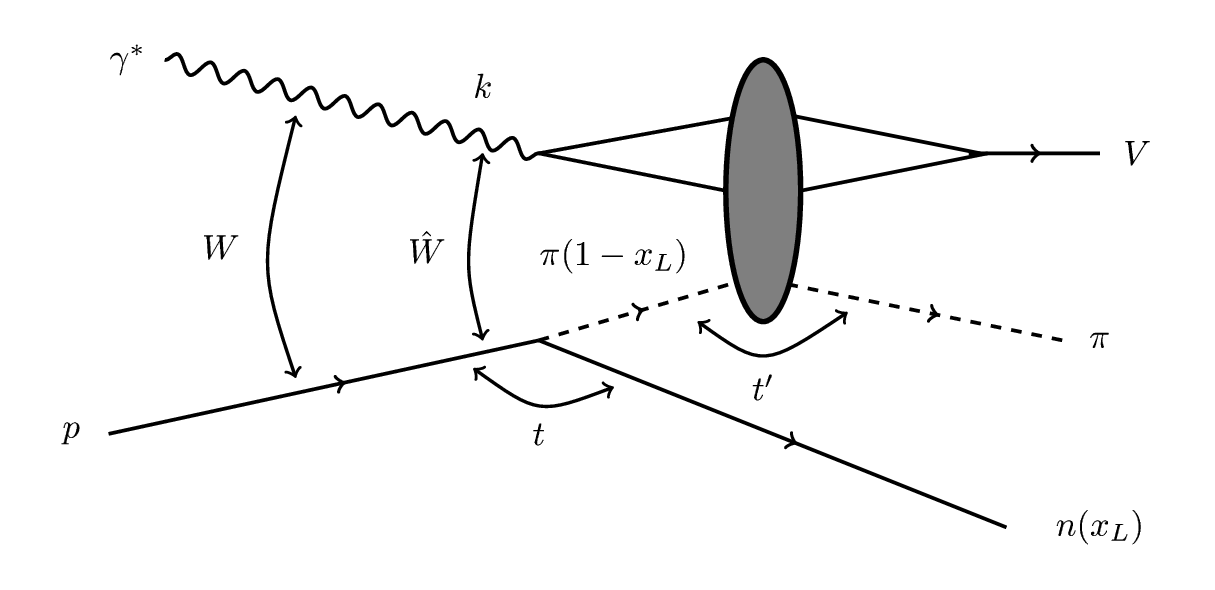}
  \caption{Exclusive $\gamma^*p$ process associated with a leading neutron $n$ production in
  the color dipole model.}
  \label{fig:process}
 \end{center}
\end{figure}

At high energies, the process can be seen as a sequence of three factorable
subprocesses, as represented in Fig.~\ref{fig:process}.
The virtual photon with four-momentum $k$, which in $ep$ collisions is emitted from the incoming
electron, splits into a quark-antiquark pair, or a dipole.
This dipole, then, interacts with a pion $\pi$ of the incident proton $p$
wave function and we have the leading neutron and the vector meson in the final
state.

In this formalism, the amplitude ${\cal A}_i^{\gamma^*\pi\to V\pi}$ reads
\begin{eqnarray}\label{eq:Axy}
{\cal A}_{T,L}^{\gamma^*\pi\to V\pi}(Y,\qq)&=&\int d^2xd^2y\int_0^1 dz
\Phi_{T,L}^{\gamma^*V}(z,\xx-\yy;Q^2,M_V^2)%\nonumber\\
%&\times &
e^{i\qq\cdot\yy}T_\pi(\xx,\yy;Y).
\end{eqnarray}
where the integration is over the quark and antiquark transverse coordinates, respectively
$\xx$ and $\yy$, and over $z$, the fraction of longitudinal momentum carried by the quark of the dipole.
The {\it overlap functions} $\Phi_{T,L}^{\gamma^*V}(z,\xx-\yy;Q^2,M_V^2)$ give the probability
of splitting of the photon into the vector meson. There are different models for them in the literature and
in this paper we will use the so-called boosted Gaussian (BG) and light-cone Gaussian (LCG) models
\cite{Nemchik,Frankfurt,Dosch} (for explicit expressions for the overlap functions, see \cite{Kowalski1,Kowalski2}).

\section{Geometric scaling and the MPS model}\label{sec:amplitude}

In Eq.(\ref{eq:Axy}), $T_{\pi}(\xx,\yy;Y)$ is the (imaginary part of the) dipole-pion
scattering amplitude. Following \cite{Carvalho:2015eia,Goncalves:2015mbf} we assume that
it can be related to the dipole-proton scattering amplitude, $T_p(\xx,\yy;Y)$ through
\begin{equation}\label{eq:Tpi-Tp}
  T_\pi(\xx,\yy;Y)=R_qT_p(\xx,\yy;Y),
\end{equation}
where the factor $R_q$ is a constant number, whose value will be discussed in the next section.
$T_p(\xx,\yy;Y)$ is the solution of BK equation
\cite{Balitsky,Kovchegov}, the simplest nonlinear evolution equation of high energy QCD,
which gives the evolution of the dipole-proton amplitude with rapidity (or, equivalently,
with energy). This equation has no exact solution, but some analytical expressions for
the amplitude can be obtained by exploring the properties of the evolution equation in
asymptotical regimes. One of these properties is that BK equation, at very high energies
(i.e., $Y\to\infty$), admits traveling wave solutions. These (analytical) solutions were
obtained for the first time in \cite{Munier}, in the simplest case where the dipole amplitude
depends only on the dipole size $r=|\rr|=|\xx-\yy|$ and the rapidity, i.e.
$T_p(\xx,\yy;Y)=T_p(r,Y)$. Besides, the analysis was done in momentum space, which can be performed
in a straightforward way through a simple Fourier transform. In this space, the dipole-proton amplitude
depends on $k=|\kk|$, where $\kk$ is the transverse momentum of the dipole. In coordinate
(momentum) space, traveling wave solutions mean that the amplitude, instead of depending on
$r$ ($k$) and $Y$ separately, depends on them through the scaling variable $rQ_s(Y)$ ($k/Q_s(Y)$),
where $Q_s(Y)$ is the saturation scale, which separates different partonic density of the proton
target, and grows exponentially with the rapidity interval, $Q^2_s(Y)= Q_0^2e^{\lambda Y}$, with
$\lambda\sim 0.3$ ($Q_0$ is a scale related to the target). This behavior of the dipole
amplitude has been a natural explanation for the observation of the so-called geometric scaling in
HERA data \cite{Stasto}, which, in the case of inclusive deep inelastic scattering, means that
the total cross section depends on the photon virtuality and energy through the variable
$Q^2/Q_s^2(Y)$.

Traveling wave solutions of BK equation can also be obtained in the case
when the full transverse dependence of the dipole amplitude is considered.
This has been shown in Refs. \cite{Marquet1,Marquet2}, and a remarkable point is that
geometric scaling is better achieved when the amplitude
depends on the dipole transverse momentum and the momentum transfer of the collision
or, in other words, is in full momentum space, and when the amplitude depends on
the dipole size and the momentum transfer, or in mixed space.
The latter case is of particular interest: after a Fourier
transform, one goes from $T(\rr,\bb;Y)$ to $\tilde{T}(\rr,\qq;Y)$, where
$\qq$ is the momentum transfer (kinematical variable conjugate to the impact
parameter). The resulting mixed-space dipole scattering
amplitude $\tilde{T}(\rr,\qq;Y)$ presents the property of geometric
scaling in the regime of small and intermediate values of the momentum
transfer $Q_0<|\qq|<Q$. In this regime, $\tilde{T}(\rr,\qq,Y)=\tilde{T}(|r|Q_s(Y),\qq)$,
with the saturation scale $Q_s^2=q^2e^{\lambda Y}$. On the other hand, in the small
momentum transfer regime, $|\qq|<Q_0<Q$, the forward result , i.e., $Q^2_s(Y)= Q_0^2e^{\lambda Y}$,
is recovered.

It is straightforward to express the amplitude (\ref{eq:Axy}) in terms of the mixed-space
dipole amplitude. First, through the change of variables $\rr=\xx-\yy$ and $\bb=z\xx+(1-z)\yy$
we rewrite it in terms of the dipole size $\rr$ and the impact parameter of the dipole-pion
interaction, $\bb$. We get
\begin{eqnarray}\label{eq:Arb}
{\cal A}_{T,L}^{\gamma^*\pi\to V\pi}(Y,\qq)&=&\int d^2rd^2b\int_0^1 dz\,
\Phi_{T,L}^{\gamma^*V}(z,\rr;Q^2,M_V^2)%\nonumber\\
%&\times&
e^{i\qq\cdot(\bb-z\rr)}T_\pi(\rr,\bb;Y)
\end{eqnarray}
Then, after introducing the Fourier transform
\begin{equation}\label{eq:Ttilde}
\tilde{T}_\pi(\rr,\qq,Y)=\int d^2b e^{i\qq\cdot b}T_\pi(\rr,\bb,Y)
\end{equation}
we finally obtain
\begin{equation}\label{eq:dsdtprime}
\frac{d\sigma_{T,L}^{\gamma^*\pi\to V\pi}}{dt^\prime}dt^\prime
=\frac{1}{16\pi}
\left|\int d^2r\int_0^1 dz\,
\Phi_{T,L}^{\gamma^*V}(z,\rr;Q^2,M_V^2)e^{-iz\qq\cdot\rr}\tilde{T}_\pi(\rr,\qq;Y)\right|^2,
\end{equation}
i.e, the cross section for the virtual photon-pion interaction in terms
of the quantity $\tilde{T}_\pi(\rr,\qq;Y)$, the dipole-pion amplitude in
mixed space.

Based on the extension of geometric scaling to the case of nonzero momentum transfer,
Marquet, Peschanski and Soyez proposed a phenomenological model for
$\tilde{T}_p(\rr,\qq;Y)$ \cite{Marquet:2007qa}. It consists in the following
expression for the mixed-space dipole-proton amplitude:
\begin{equation}\label{eq:MPS}
\tilde{T}_p(\rr,\qq;Y)=2\pi R_p^2f(\qq)N(rQ_s(Y,\qq)),
\end{equation}
where $R_p$ is the proton radius, $f(\qq)$ is a form factor, which
catches the transfer dependence of the proton vertex, given by
$f(\qq)=\exp^{-B\qq^2}$. The function $N$ is a generalization of the
Iancu, Itakura and Munier (IIM) forward saturation model \cite{Iancu:2003ge}:
\begin{equation}\label{eq:iim}
N(rQ_s(Y),Y)=\left\{\begin{array}{ll}
N_0\frac{rQ_s(Y)}{2}^{2\left(\gamma_c+\frac{\ln(2/rQ_s)}{\kappa\lambda Y}\right)},& rQ_s(Y) \leq 2;\\
   1-e^{-a\ln^2(brQ_s(Y))}, & rQ_s(Y) > 2,
\end{array}
\right.
\end{equation}
with a saturation scale depending on the momentum transfer, parametrized as
\begin{equation}\label{eq:Qs}
Q_s^2(Y,\qq) = Q_0^2(1+c\qq^2)e^{\lambda Y}.
\end{equation}
This is a simple and intuitive phenomenological model which completely recovers the
geometric scaling in the forward case ($\qq=0$) and gives the saturation
scale the correct asymptotic behaviors. Using the MPS model, Eqs. \eqref{eq:MPS}-\eqref{eq:Qs},
and the relation \eqref{eq:Tpi-Tp} we are able to calculate the cross section
\eqref{d2sdxdt} for the exclusive vector meson production with a leading neutron
and confront to available data.

%%%%%%%%%%%%%%%%%%%%%%%%%%%%%%%%%%%%%%%%%%%%%%%%%%%%%%%%%%%%%%%%%
\section{Results}\label{sec:results}

In this analysis we consider the HERA data on exclusive photoproduction
of $\rho^0$ mesons, associated with leading neutrons, presented in \cite{Andreev2016}.
The data covers the following kinematical ranges: $20<W<100$ GeV and $Q^2<2$ GeV$^2$.
The range on the $Q^2$ leads to a mean value $\langle Q^2 \rangle=0.04$ GeV$^2$.
Following \cite{Goncalves:2015mbf}, here we use $W=60$ GeV and $Q^2=0.04$ GeV$^2$.
All the parameters of MPS model are kept fixed. Those who enter the scaling function
$N$ are taken from Ref.\cite{Soyez:2007kg}, where the IIM saturation model was extended
to include heavy quarks, the values of their masses being $m_f=0.14$ GeV for light flavors,
$m_c=1.4$ GeV for charm quark and $m_b=4.5$ GeV for bottom quark.
The parameters $a$ and $b$ are uniquely determined from the
conditions that $N$ and its derivative are continuous at $rQ_s=4$ and $N_0$ is fixed at
the value $N_0=0.7$. The saturation scale parameters are $\lambda=0.2197$ and
$Q_0 = 0.298$ GeV and the proton radius is $R_p=3.34$ GeV$^{-1}$. The remaining parameters
of MPS model are $c$ and $B$ and their values are chosen to be those which provided the best
description of the HERA data on exclusive vector meson production in \cite{Marquet:2007qa},
$c=4.401$ and $B=3.713$, respectively.

Another parameter which enters the calculations is $R_q$, which relates the dipole-proton and dipole-pion
scattering amplitude, Eq.(\ref{eq:Tpi-Tp}). Its value, according to the additive quark model,
is expected to be 2/3, the ratio between the number of valence quarks in both targets. This was
the same value obtained for $R_{\pi/ p}\equiv \sigma_{\pi p}/\sigma_{pp}$, the ratio between
the pion-proton and proton-proton cross sections, when the quark model was applied to soft hadronic
interactions \cite{Levin:1965mi}. $R_{\pi/ p}=2/3$ was also experimentally observed in the low energy
domain of hadronic reactions, provided good description of previous ZEUS data on leading neutron
spectra \cite{Kaidalov2006,Khoze2006} and is supported by the investigation of the pion structure function
done in \cite{Nikolaev2000157}. On the other hand, when relation (\ref{d2sdxdt})
was applied to HERA data on photoproduction \cite{Chekanov2002}, the resulting value for this ratio
was $R_{\pi p}\simeq 1/3$ and, in recent work, it has been concluded that it could reach 0.5
\cite{PhysRevD.85.114025}. Therefore, as we can see, the value for the parameter $R_q$ is
still an open question, and what we can say is that it is expected to be in the range $1/2 \leq R_q \leq 2/3$.
Aiming at a comparison with the analysis done in coordinate space in \cite{Goncalves:2015mbf},
in the present analysis we will keep it fixed at $R_q=2/3$.

The cross section for leading neutron production \eqref{d2sdxdt} depends on the
pion flux, whose generic form reads
\begin{equation}\label{eg:general-pion-flux}
f_{\pi/p}=\frac{1}{4\pi}\frac{2g_{p\pi n}^2}{4\pi}
\frac{-t}{(t-m_\pi^2)^2}(1-x_L)^{1-2\alpha(t)}
\left[F(x_L,t)\right]^2,
\end{equation}
where $g^2_{p\pi n}/4\pi=14.11$ is the $p\pi n$ coupling constant \cite{Ericson:2000md},
$\alpha(t)$ is the
pion trajectory and $F(t,x_L)$ is a model-dependent form factor, which accounts
for the finite size of the nucleon and the pion. As in \cite{Goncalves:2015mbf},
here we consider five parametrizations for the form factor:
\begin{equation}
 \label{f_1}
 F_{1}(x_{L},t)=\exp\left[R^{2}\frac{(t-m_{\pi}^{2})}{(1-x_{L})}\right],\quad \alpha(t)=0
\end{equation}
\cite{HOLTMANN1994363}, where $R=0.6$ GeV$^{-1}$;
\begin{equation}
 \label{f_2}
 F_{2}(x_{L},t)=1,\quad \alpha(t)=\alpha(t)_{\pi}
\end{equation}
\cite{BISHARI1972510}, where $\alpha(t)_{\pi}\simeq t$ ($t$ is given in GeV$^{2}$);
\begin{equation}
 \label{f_3}
 F_{3}(x_{L},t)=\exp\left[b(t-m_{\pi}^{2})\right],\quad \alpha(t)=\alpha(t)_{\pi}
\end{equation}
\cite{Kopeliovich1997}, where $\alpha(t)_{\pi}\simeq t$ ($t$ is given in GeV$^{2}$)
and $b=0.3$ GeV$^{-2}$;
\begin{equation}
 \label{f_4}
 F_{4}(x_{L},t)=\frac{\Lambda_{m}^{2}-m_{\pi}^{2}}{\Lambda_{m}^{2}-t},\quad \alpha(t)=0
\end{equation}
\cite{PhysRevD.43.59}, where $\Lambda_{m}=0.74$ GeV;
\begin{equation}
 \label{f_5}
 F_{5}(x_{L},t)=\left[\frac{\Lambda_{d}^{2}-m_{\pi}^{2}}{\Lambda_{d}^{2}-t}\right]^{2},\quad \alpha(t)=0
\end{equation}
\cite{PhysRevD.43.59}, where $\Lambda_{d}=1.2$ GeV.
For each of these expressions for the form factor one has a corresponding model for the pion flux,
which we will call, respectively, $f_1$, $f_2$, $f_3$, $f_4$ and $f_5$.

Finally, a crucial point of the analysis of processes involving leading neutron production
is that the absorptive corrections must be taken into account. These arise because
the photon eventually also hits  the neutron, leading to extra interactions and,
therefore to a reduction of the cross section. Usually, these absorptive corrections
are represented by a constant factor $K$, which multiplies the (uncorrected) cross
section \eqref{d2sdxdt}. Following the analysis done in \cite{Goncalves:2015mbf},
we obtain the $K$-factor through the ratio between the experimental and theoretical
(calculated) total cross sections. The uncertainty on the former translates into a range
of possible values for $K$, which will be around a central value $K_{\textrm{med}}$, between a
minimum value $K_{\textrm{min}}$ and a maximum value $K_{\textrm{max}}$.

\begin{table*}[h!]
\caption{$\chi^2/N_{\textrm{pts}}$ for all the different combinations of the models for
the overlap functions and the pion flux $f_{\pi/p}$.}
\label{tab:chi2}
\begin{center}
\begin{tabular}{cccccc}
\toprule
{} & \multicolumn{5}{c}{$\chi^2/N_{\textrm{pts}}$ }\\
\cmidrule(r){2-6}
{Model for overlap functions} &\ {$f_1$} & {$f_2$} & {$f_3$} & {$f_4$} & {$f_5$}\\
\midrule
BG   &{\bf 1.178}  &0.724  &$0.559$  & 1.462  &1.559  \\
LCG   &{\bf 1.170}  &0.718  &0.559 &  1.468  &1.564 \\ \hline
\bottomrule
\end{tabular}
\end{center}
\end{table*}

\begin{figure*}
 \begin{center}
  \includegraphics[scale=0.51]{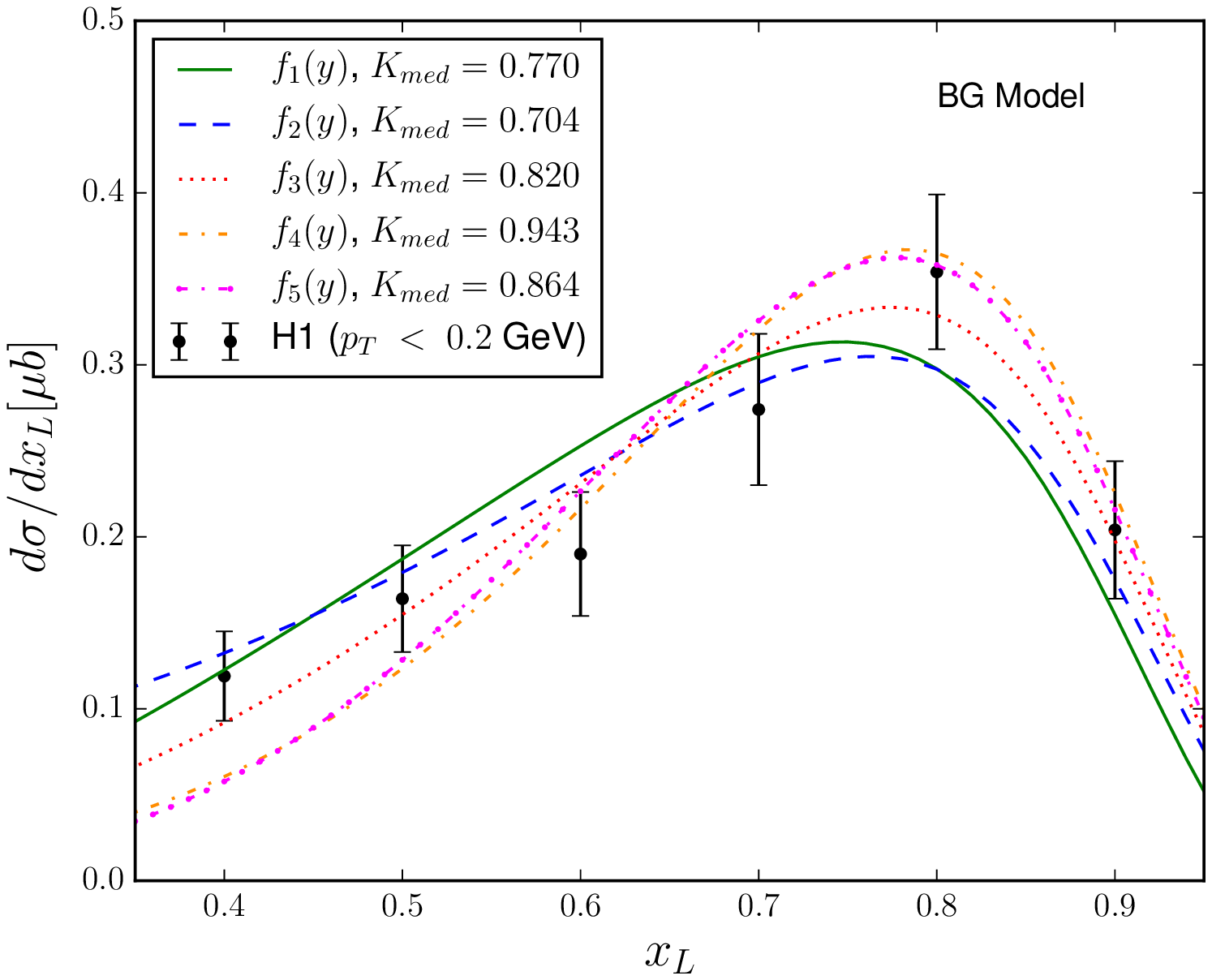}
  \includegraphics[scale=0.51]{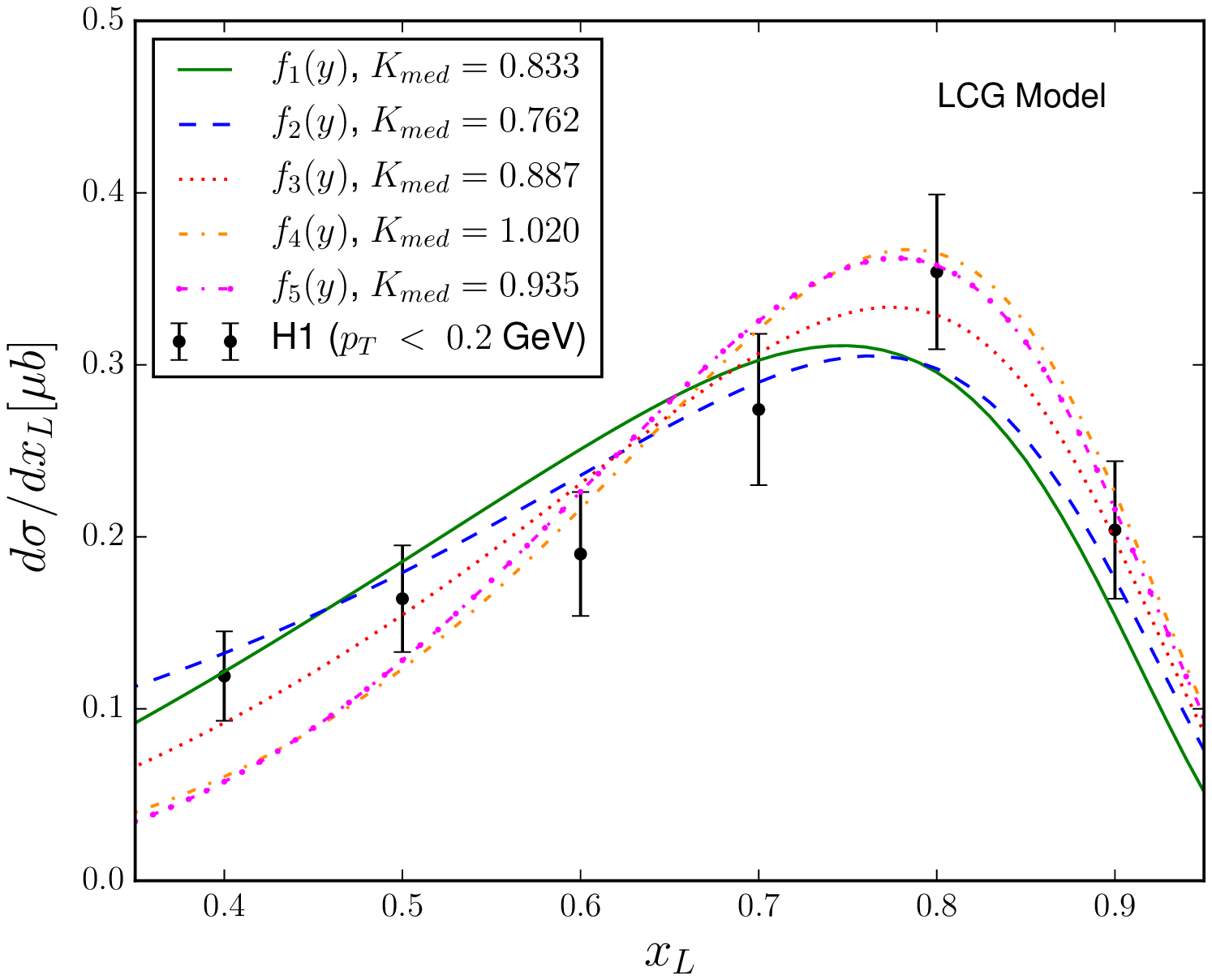}
  \caption{Leading neutron spectra in exclusive $\rho$ production considering the boosted gaussian (left)
  and the light-cone Gauss (right) models for the vector meson wave function for the five different models of the pion flux. The data were taken from \cite{Andreev2016}.}
  \label{fig:dsdxL-all}
 \end{center}
\end{figure*}

Our first results are presented in Fig.~\ref{fig:dsdxL-all}, which shows the differential cross section
$\frac{d\sigma(W,Q^2,x_L)}{dx_L}$ as a function of $x_L$, where the range in the leading neutron transverse
momentum is $p_{T,n}<0.2$ GeV, one of the kinematical ranges covered by the data in \cite{Andreev2016}.
We consider the five models for the pion flux presented above and both BG and LCG models for the
overlap functions. The value of the resulting $K$-factor in each case is also given. 
As in the
coordinate-space analysis, for all the models for the pion flux, the LCG model for the overlap
functions tends to provide larger values of $K$, in comparison with BG model. 
We can also observe
that, as in the coordinate-space case, model $f_2$ for the pion flux gives the smallest value for
$K$, while $f_4$ gives the largest one. Among the five models for the pion flux, the one which provides the best description of the data is
$f_1$, as we can see in Table \ref{tab:chi2}, where we present the calculated $\chi^2$ over the number
of points for all the combinations of models for the pion flux and overlap functions. This is also in
contrast with the coordinate-space study, in which the best descriptions of the data were obtained
by using models $f_2$ and $f_3$. From Table \ref{tab:chi2} we can also see that, in which concerns the goodness
of the fit to the data, there is no preference for on of the models for the overlap functions.

\begin{figure*}
 \begin{center}
  \includegraphics[scale=0.5]{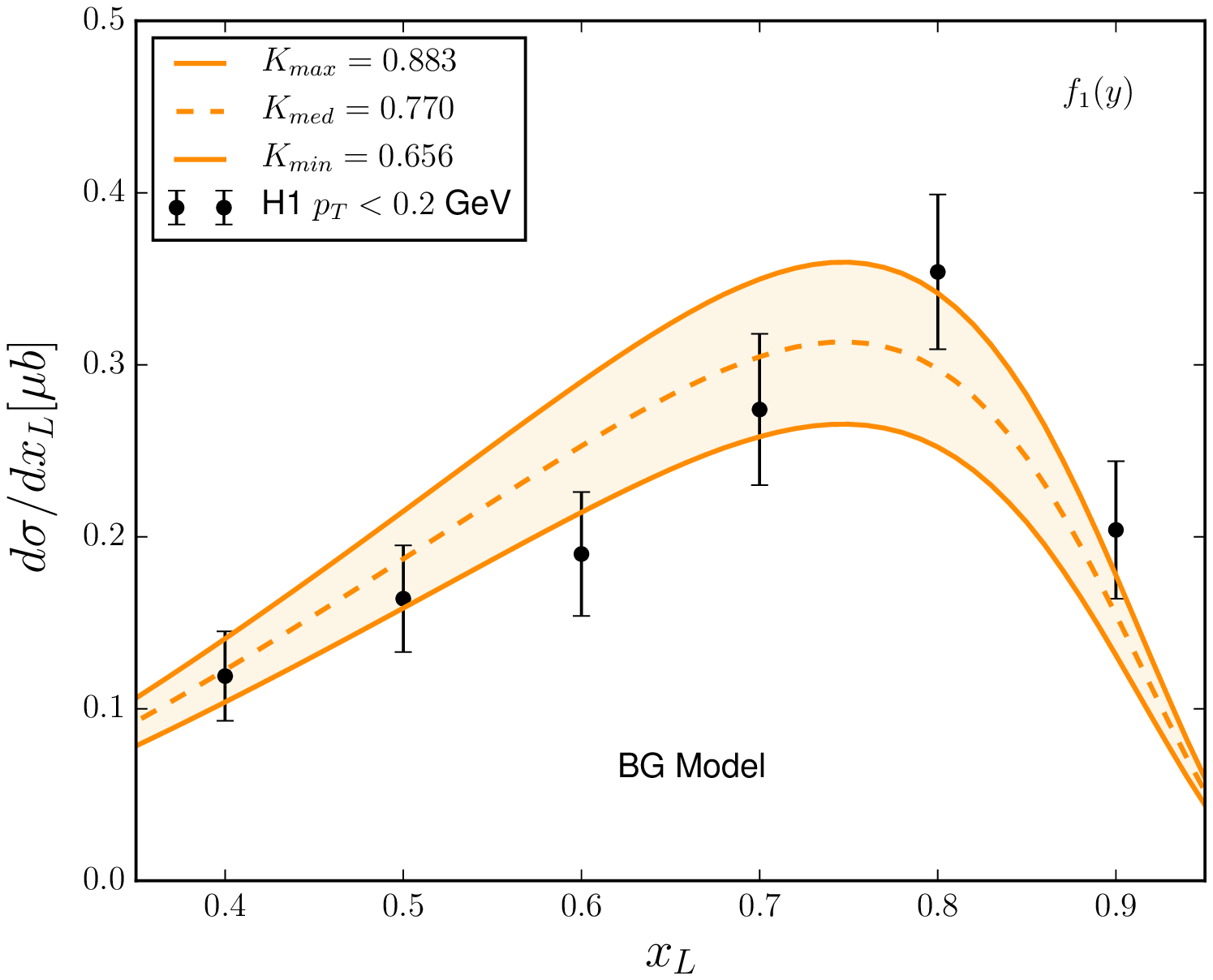}
  \includegraphics[scale=0.5]{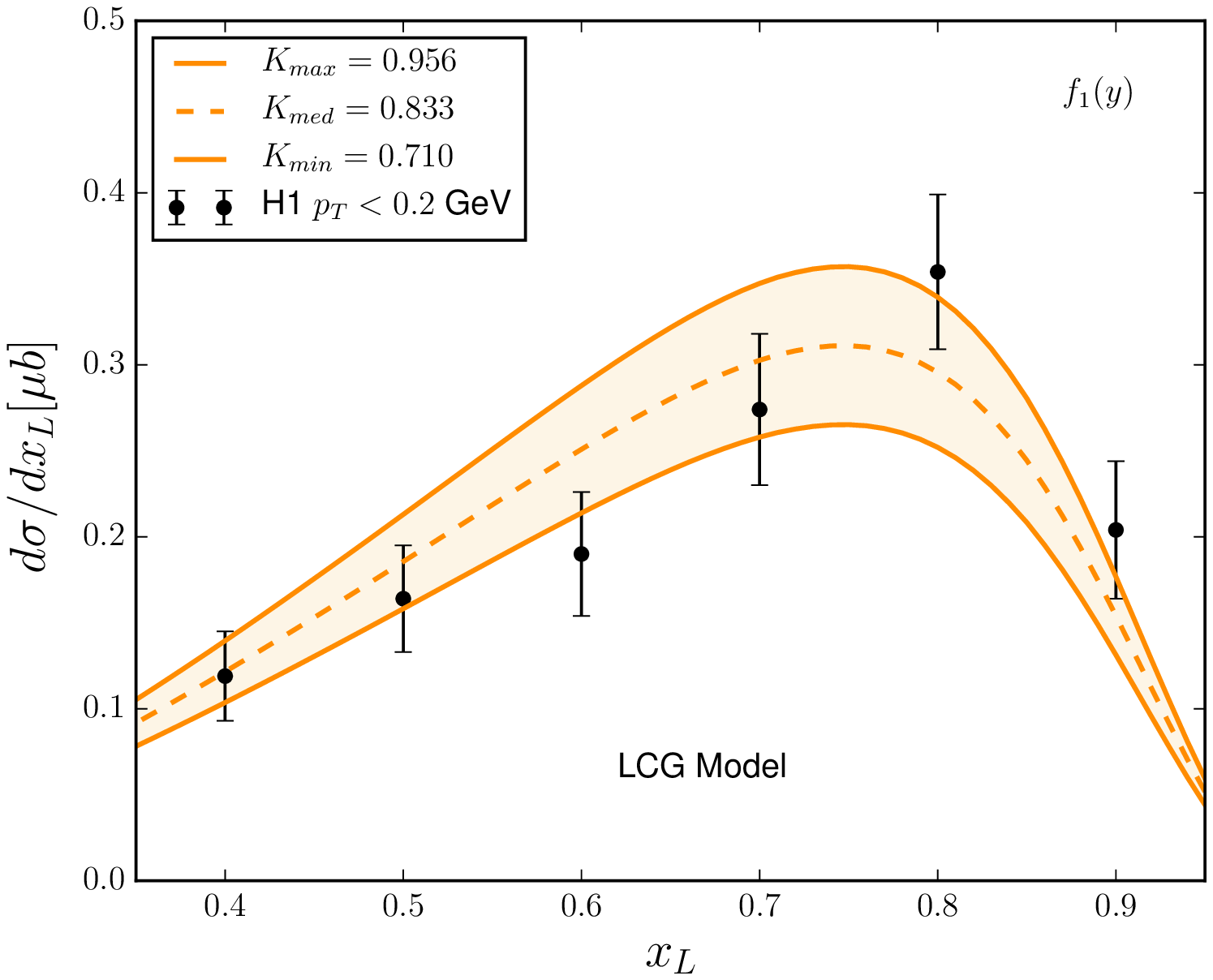}
  \caption{Leading neutron spectra in exclusive $\rho$ production for the pion flux model $f_{1}$ considering
  the boosted Gaussian (left) and the light-cone Gauss (right) models for the vector meson wave function.
  The possible range of values of the $K$ factor are shown.
  The data were taken from \cite{Andreev2016}.}
  \label{fig:f1-pt02}
 \end{center}
\end{figure*}

The most remarkable difference between our approach and the coordinate space one is
that the magnitude of the absorption corrections are quite different. These corrections, mimicked
by an overall constant $K$ factor, are related to rescattering of the projectile photon. As
in \cite{Goncalves:2015mbf}, we estimated its range of values by taking into account, as a constraint, the experimental value for the total cross section of the process \cite{Andreev2016}. 
The results are presented in Fig. \ref{fig:f1-pt02}, where we show our results using the model $f_1$
for the pion flux, including the possible range of values for the $K$-factor, with both BG and
LCG models for the overlap functions. This range is $K\simge 0.65$ to $K\simle 0.89$, when BG model is used for the overlap functions, and $K\simge 0.70$ to $K\simle 0.96$ when LCG model is used.
This is in a strong contrast to the results obtained in \cite{Goncalves:2015mbf}, where it was found that
the $K$-factor falls in a quite different range, going from $K\simge 0.11$ to $K\simle 0.16$ when using
$f_2$ and $K\simge 0.15$ to $K\simle 0.21$ when using $f_3$ for the pion flux. Thus,
in comparison with the coordinate study using bCGC model, the present analysis, using the mixed-space
MPS model, indicates that absorption effects are substantially weaker and of the same order of the
corresponding corrections in semi-inclusive processes.

\begin{figure*}
 \begin{center}
  \includegraphics[scale=0.5]{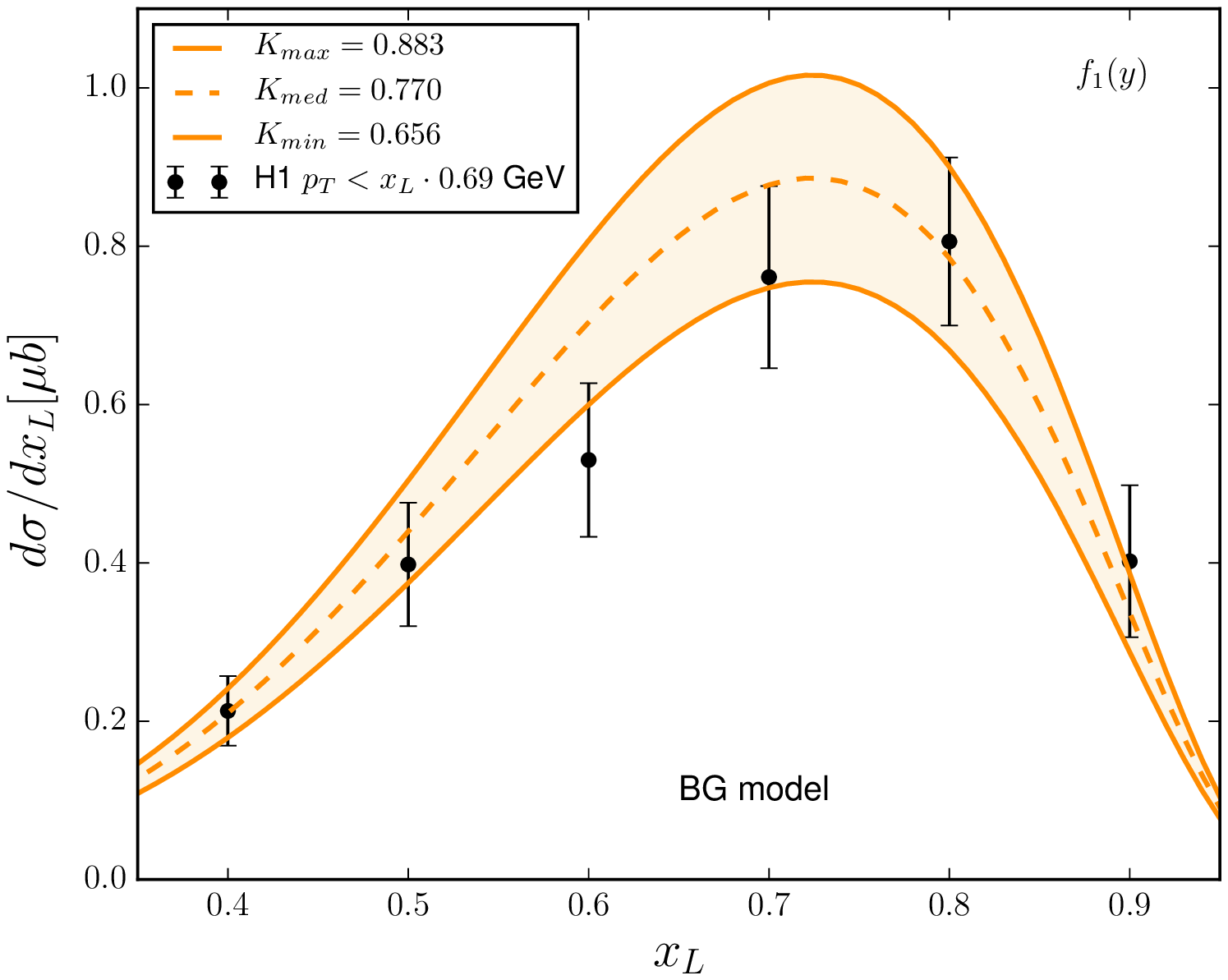}
  \includegraphics[scale=0.5]{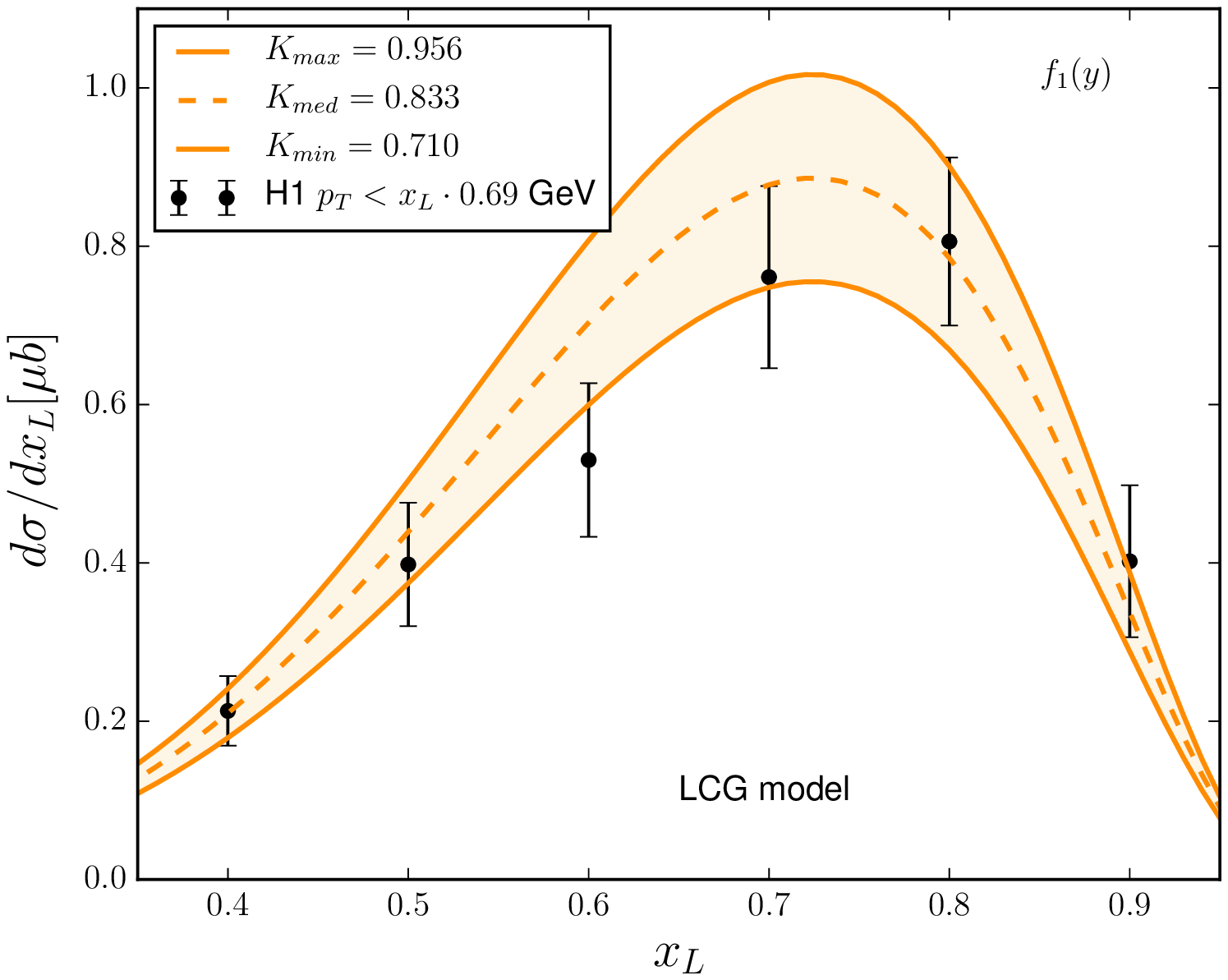}
  \caption{Prediction for the leading neutron spectra in exclusive $\rho$ production considering $p_{T}<0.69\cdot x_{L}$ GeV.
  The possible range of values of the $K$ factor for the pion flux model $f_{1}$ is fixed by the previous set
  of experimental data.
  The data were taken from \cite{Andreev2016}.}
  \label{fig:f1-pt069xL}
 \end{center}
\end{figure*}

We also compared our predictions to other H1 data in a different
range of the transverse momentum of the leading neutron, $p_{T,n}<0.69x_L$ GeV. Using the model $f_1$
for the pion flux and following \cite{Goncalves:2015mbf}, we fixed the same ranges for $K$ obtained in
Fig. \ref{fig:f1-pt02} and use them to describe these data. The results are shown in Fig.
\ref{fig:f1-pt069xL}, for both BG and LCG models for the overlap functions. The same procedure
was done using all the other models for $f_{\pi/p}$ and we obtained, again, that $f_1$ provides
the best description of the data, again with no preference between BG and LCG models for the overlap functions.

%%%%%%%%%%%%%%%%%%%%%%%%%%%%%%%%%%%%%%%%%%%%%%%%%%%%%%%%%%%%%%%%%%%%%%%%%%%%%%%%%%%%%%%%%%

\section{Conclusions}\label{sec:conclusions}

In this work we used a mixed-space approach of the color dipole formalism to study
the exclusive vector meson production in $ep$ collisions involving leading neutron. 
For the mixed-space dipole-pion scattering amplitude we used the MPS phenomenological 
model for the dipole-proton scattering amplitude, which is based on the traveling wave
solutions of BK equation, the simplest nonlinear evolution equation of high energy QCD.
Following the full coordinate-space analysis done in \cite{Goncalves:2015mbf}, we
assumed a simple relation between the dipole-pion and dipole-proton amplitudes,
Eq.(\ref{eq:Tpi-Tp}) with fixed $R_q=2/3$, tested
five models for the pion flux and confronted our results to recent HERA data
on exclusive $\rho$ meson production with leading neutron.
From the five models used for the pion flux, Eqs.(\ref{f_1})-(\ref{f_5}),
model $f_1$ for the pion flux provided the best description
of the data, in contrast to the full coordinate-space analysis, in which the best
descriptions were obtained by using models $f_2$ and $f_3$ Eqs.(\ref{f_2}) and (\ref{f_3}).

We also estimated the magnitude of the absorption effects, usually included in an overal $K$-factor,
and our results indicate that in exclusive processes involving leading neutrons
the absorption effects are weaker in comparison with the full coordinate-space approach.
This can be verified by the possible values for the $K$-factor, which according to our
analysis should not be smaller than $0.6$ (being of the order of those predicted for semi-inclusive
processes), while in the coordinate-space one they should not be larger than $0.3$.
It should be pointed out that here, as in \cite{Goncalves:2015mbf}, the values obtained
for $K$ were directly related to the choice for the parameter $R_q$, which in the coordinate
approach translates into a large flexibility in the values of the former as the latter goes from
1/3 to 2/3. This is not the case in the present analysis, which clearly indicates that $R_q$
could be smaller than, but not so far from, the value predicted by the additive quark model.

From the above conclusions we can see that, as long as more data and theoretical studies
on exclusive processes with leading neutron become available, we will be able to discriminate
between both (coordinate and mixed space) approaches and, therefore, shed new light on
(or even reduce) other uncertainties involving these processes, i.e., the more suitable model for the pion flux,
absorption corrections and the relation between the dipole-pion and dipole-proton scattering amplitudes.

%%%%%%%%%%%%%%%%%%%%%%%%%%%%%%%%%%%%%%%%%%%%%%%%%%%%%%%%%%%%%%%%%%%%%%%%%%%%%%%%%%%%%%%%%%

\section*{ACKNOWLEDGMENTS}
We would like to thank Victor Gon\c{c}alves and Diego Spiering for very useful discussions.
This work was partially supported by CAPES (Brazilian research funding agency).

%%%%%%%%%%%%%%% BIBLIOGRAPHY %%%%%%%%%%%%%%

\end{document}